\documentstyle[amssymb,11pt,epsf]{article}
\begin{document}

\title{{\bf Single top production at the LHC:}\\
{\bf the Effective $W$ Approximation\thanks{%
Contribution presented in the XXIX International Meeting on Fundamental
Physics, Sitges, February 2001. Dedicated to F.J.Yndur\'{a}in on his 60th
birthday.}}}
\author{D. Espriu\thanks{%
espriu@ecm.ub.es} \ and J. Manzano\thanks{%
manzano@ecm.ub.es} \\
\\
%EndAName
Departament d'Estructura i Constituents \\
de la Mat\`{e}ria and IFAE,\\
Universitat de Barcelona, \\
Diagonal, 647, E-08028 Barcelona}
\date{}
\maketitle

\begin{abstract}
Motivated by the need to set bounds on the third generation charged
couplings, we study the mechanism of single top production at the LHC,
analyzing the sensitivity of different observables to the magnitude of the
effective left and right couplings. The study is carried out in the
framework of the so-called effective W approximation, where the virtual $W$
is treated as a parton. We take this opportunity to critically assess the
validity of this approximation in detail by comparing it to an exact
calculation recently completed by us of the same process. We comment on
several issues related to top polarization since the observables relevant to
distinguish between left and right effective couplings involve the
measurement of the spin of the top. The conclusion is that the effective $W$
approximation is not well suited for this subtle analysis; it fails to
reproduce the detailed $p_T$ distributions and grossly distorts the
polarization of the emerging top. It does however, reproduce the overall
angular distribution and gives a sensible order-of-magnitude estimate for
the total cross section analysis.
\end{abstract}

%\title{{\bf Single 
%top production
%at the LHC: the Effective $W$  Approximation\footnote{%
%Contribution presented in the XXIX International Meeting on 
%Fundamental Physics, Sitges, February 2001. Dedicated to F.J.Yndur\'ain on
%his 60th birthday.}}}
%\author{{{\sc D. Espriu}\footnote{%
%espriu@ecm.ub.es.}}\ and \ {{\sc J. Manzano}\footnote{%
%manzano@ecm.ub.es}} \\
%{Departament d'Estructura i Constituents de la Mat\`eria and IFAE} \\
%{\ Universitat de Barcelona, Diagonal, 647, E-08028 Barcelona}
%\\
%\\
%}
%\date{}
%\maketitle

\vfill
\vbox{
UB-ECM-PF 01/09\null\par
August 2001\null\par
}

\clearpage

\section{Introduction}

The main purpose of this work is to analyze the applicability of the
so-called effective-$W$ approximation (to be described below) to the study
of single top production at the LHC, the aim being to set bounds on the
effective couplings for the charged couplings involving the third generation.

It is quite conceivable that the Standard Model should be considered as an
effective theory valid only at low energies ($\lesssim 1$ TeV ). In
particular, since the Higgs particle has not been observed yet (the current
bound on the Standard Model Higgs is at 113.5 GeV \cite{LEP}), it makes
sense to consider as an alternative to the minimal Standard Model an
effective theory without any physical light scalar fields. Alternatively, it
may well be that the Higgs particle, or, as a matter of fact, any other
scalars, which abound in extensions of the minimal Standard Model, are much
heavier than the weak scale, granting an expansion in inverse powers of such
heavy masses. How heavy must the Higgs particle ---or the scale associated
to new physics, for that matter--- be for such an expansion to be useful? In
practice, 300 or 400 GeV are sufficient for the expansion to be useful at
the $M_{Z}$ scale, as detailed calculations\cite{DEH} show. In a process
like the one we shall discuss in this paper, where the individual energies
involved are typically peaked in the 200 - 400 region, the momentum
expansion should be appropriate provided that the relevant scale for the new
physics is in the 1 TeV region or beyond. While admittedly this is not the
scenario favoured by the comparison with the electroweak data\cite
{PeskinWells}, it cannot be properly excluded until a (relatively) light
elementary Higgs is found.

In the effective lagrangian, only the light (respect to the energies of the
process) degrees of freedom are kept, while the information about the
heavier degrees of freedom is contained in an infinite set of effective
operators of increasing dimensionality, compatible with the electroweak and
strong symmetries $SU(3)_{c}\times SU(2)_{L}\times U\left( 1\right) _{Y}$.
The coefficients of these operators would parametrize different choices of
new physics beyond the Standard Model. In this framework \cite{eff-lag} one
can describe the low energy physics of theories exhibiting the pattern of
symmetry breaking $SU(2)_{L}\times U\left( 1\right) _{Y}\rightarrow U\left(
1\right) _{em}$. Both global and gauge symmetries are non-linearly realized
and the effective theory is non-renormalizable (the Higgs field, which is
absent here, is a necessary ingredient both for the linear realization and
renormalizability of the minimal Standard Model). The additional operators
serve thus a dual purpose; on the one hand they encode low-energy effects of
the so-far unexplored high-energy scales. On the other hand, these operators
are necessary as counterterms to absorb ultraviolet divergences generated by
quantum corrections from the lower dimensional (universal) terms.

In this work we are concerned about the new features that physics beyond the
Standard Model may introduce in the production of single top quarks through $%
W$-gluon fusion at the LHC. We are interested only in the leading
non-universal (i.e. not appearing in the standard model at tree level)
effective operators in the low energy expansion. In the present context
these correspond to those operators of dimension four, which were first
classified by Appelquist et al. \cite{appel}. These operators are
characteristic of strongly coupled theories (i.e. of theories without an
elementary Higgs or a very heavy one) and require a non-linear realization
of the gauge symmetry. Therefore they are absent in the minimal Standard
Model and in modifications thereof containing only light fields. When one
particularizes to the $W$ interactions, for instance by going to the unitary
gauge, these operators induce effective fermion-gauge boson couplings, and
these effective couplings are the object of our interest.

Of course radiative corrections induce form factors in the vertices too.
Assuming a smooth dependence in the external momenta these form factors can
be expanded and at leading order in the derivative expansion they just
induce effective fermion-gauge boson effective couplings, exactly as the
putative contribution from new physics. These radiative corrections are
typically very small, at the few per cent level, but in general
non-negligible (particularly those involving the top mass). Obviously any
deviation from the values of these couplings with respect to the values
predicted by the minimal Standard Model would indicate the presence of new
physics in the matter sector. The extent to which a machine like the LHC can
set direct bounds on these couplings, in particular on those involving the
third generation, is thus of obvious interest.

In fact there if anywhere are deviations with respect to the Standard Model
to show up in the matter sector, this is the place to look for them. Many
alternatives to the minimal Standard Model (dynamical symmetry breaking
models, for instance) predict large deviations for the third generation
effective couplings in a natural way (typically much larger than radiative
corrections from the minimal Standard Model itself). The fact that the
longitudinal degrees of freedom of the vector bosons ---the very product of
the electroweak symmetry breaking---have couplings that, after use of the
equations of motion, are proportional to the quark masses hint in this same
direction too. For all these reasons we regard the possibility of getting a
handle on such effective couplings as one of the main tasks of the LHC
experiments.

\section{Top Production at the LHC}

Let us now briefly review the mechanisms of top production. At the LHC
energy (14 TeV) the dominant mechanism for creating tops is gluon-gluon
fusion. This is a purely QCD process, its total cross-section is 800 pb\cite
{catani}. It is obvious that tops will be copiously produced at the LHC and
that a lot can be learned by a detailed analysis of their decays, for
instance. We note, however, that this mechanism of production has nothing to
do with the electroweak sector and thus is not the most adequate for our
purposes. In fact, since colour symmetry, like electromagnetism, is
unbroken, the form factor is unchanged at the lowest order in the derivative
expansion and new physics cannot possibly affect this effective coupling at
the order we are working \cite{EspManzCP}. In addition, we shall not be
interested here in top decay, but rather in how new physics can affect the
way top (or anti-top) are produced at the LHC. For these reasons, the
dominant mechanism of top production at the LHC is not interesting at all
for our purposes, and for us it will just be a background to worry about.

\begin{figure}[tbp]
\epsfysize=2.7cm %\epsfxsize=10.0cm
\centerline{\epsffile{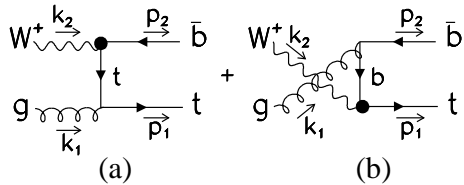}}
\caption{Feynman diagrams contributing to the single top production
subprocess}
\label{t+b+}
\end{figure}

At tree level, electroweak physics enters the game through single top
production. (For a recent review see e.g. \cite{Tait}.) At LHC energies the
(by far) dominant electroweak subprocess contributing to single top
production is given by a gluon ($g$) coming from one proton and a positively
charged $W^{+}$ coming from the other (this process is also called $t$%
-channel production\cite{tchannel,SSW}). This process is depicted in
diagrams (a) and (b) of Fig.\ref{t+b+}. The total cross section for this
process at the LHC is 250 pb, to be compared to 50 pb for the associated
production\footnote{%
the difference between the two processes is purely kinematic; see section 
\ref{cross}. The above values correspond to the cuts used in \cite{SSW}}
with a $W^{+}$ boson and a $b$-quark extracted from the sea of the proton,
and the 10 pb corresponding to quark-quark fusion ($s$-channel production).
For a detailed discussion see \cite{SSW}. For comparison, at the Tevatron (2
GeV) the cross section for $W$-gluon fusion is 2.5 pb, so the production of
tops through this particular subprocess is really copious at the LHC. Monte
Carlo simulations including the analysis of the top decay products indicate
that this process can be analyzed in detail at the LHC and traditionally has
been regarded as the most important one for our purposes.

\begin{figure}[tbp]
\epsfysize=2.7cm %\epsfxsize=10.0cm
\centerline{\epsffile{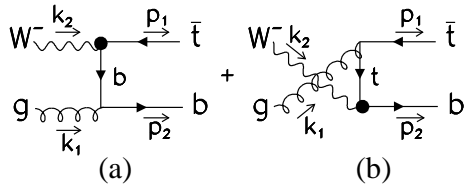}}
\caption{Feynman diagrams contributing to the single anti-top production
subprocess}
\label{t-b-}
\end{figure}

In a proton-proton collision a bottom-anti-top pair is also produced,
through the subprocesses (a) and (b) of Fig.\ref{t-b-}. However, these
subprocesses are suppressed roughly by a factor of two (see Fig.\ref
{ptatopeffw_lab_lr_gl=1_gr=0}) because the proton has much lesser
probability of emitting a $W^{-}$ than emitting a $W^{+}$, and at any rate
qualitative results are very similar to those corresponding to the
subprocess of Fig.\ref{t+b+}, from where the cross sections can be easily
derived doing the appropriate changes.

Thus, even if subdominant, single top production through an electroweak
vertex is not negligible at all; more than one third of all tops or
anti-tops that will be produced at the LHC will be created through this
mechanism. Furthermore, this mechanism has rather distinctive kinematics, as
we shall see in this work. Indeed, single top production, being dominated by
the exchange of a massless (by comparison to the energies involved) particle
in the $t$ channel, is strongly peaked in the beam direction, with a
characteristic angular distribution (see below).

Several analysis of top production exist in the literature. A (surely
incomplete) list of the references we have used is given in \cite
{Tait,tchannel,SSW} and \cite{parke,Mauser}. The second group of references
is mostly concerned with the issue of the top polarization. Indeed, since
the top decays shortly after production, much before strong effects can set
in, the decaying products (a $b$ quark and a $W$, which, in turn, decays
into e.g. a charged lepton and a neutrino) carry information about the spin
of the top. In particular, if the top is in a pure state of spin (and hence
being a $s=1/2$ fermion its polarization vector points in the top rest frame
in a particular direction in space), the decaying lepton has an angular
distribution peaked in the direction where the top polarization vector is
pointing to.

So far the issue as to what extend LHC can set bounds on the top-bottom
effective couplings has not been analyzed in much detail in the literature.
In the effective lagrangian language, the contribution from operators of
dimension five to top production via longitudinal vector boson fusion was
estimated some time ago in \cite{LY}, although the study was by no means
complete. It should be mentioned that $t,\bar{t}$ pair production through
this mechanism is very much masked by the dominant mechanism of gluon-gluon
fusion, while single top production, through $WZ$ fusion, is expected to be
quite suppressed compared to the mechanism presented in this paper, the
reason being that both vertices are electroweak in the process discussed in 
\cite{LY}, and that operators of dimension five are expected to be
suppressed, at least at moderate energies, such as the ones that in practice
count at LHC, by some large mass scale. The effects due to anomalous
couplings introduced through operators of dimension six have been recently
analyzed in \cite{zhang2}. All the contributions from higher dimensional
operators are expected in realistic models to be small and far beyond the
sensitivity of LHC. For these reasons we concentrate only on dimension four
operators. In practice this means bounds on the left and right top effective
couplings (see also \cite{zhang}) thus providing further reasons for a good
measurement of these parameters.

The potential for single top production for measuring the CKM matrix element 
$V_{tb}$, and hence the top-bottom effective left coupling has certainly
been previously studied (see e.g. \cite{tchannel}, \cite{SSW} and our recent
work \cite{exact}), but drawing firm conclusions requires a good knowledge
of the total normalization of the cross-section, something which is very
influenced by issues on which a good theoretical control is problematic,
such as the QCD scale. Next to leading calculations such as the ones
presented in \cite{SSW2,SmW} for the total cross section are mandatory but
results still leave an uncertainty at the 5\% level, so this will be the
final level of uncertainty on the left effective coupling. Nothing is known
on bounds on the right effective couplings. On the other hand, it is clear
that this is a very urgent issue. The only noticeable discrepancy of the
whole of the LEP results when compared to the minimal Standard Model lies in
the bottom effective couplings (actually on the right coupling). Our
original motivation for this work was to partly fill in this gap. This looks
feasible with machines like the LHC which are efficient top factories.

\section{Effective Couplings}

The alert reader may have noticed in the previous paragraphs that we attach
a lot of importance to the contribution of dimension four operators ---they
may encode the larger contribution in a large family of extensions of the
Standard Model.

On the other hand, is sometimes stated that gauge invariance prevents the
presence of dimension four operators other than the ones already existing in
the Standard Model, thus forcing the contribution of new physics to be
suppressed by powers or otherwise renormalizing the gauge coupling constant.
This widespread belief is not correct when the symmetry is non-linearly
realized. The complete set of dimension four effective operators (which may
eventually contribute to the top effective couplings) is \cite{appel,Bagan} 
\[
\begin{array}{lcl}
{\cal L}_{4}^{1}=i\delta _{1}{\rm \bar{f}}\gamma ^{\mu }U\left( D_{\mu
}U\right) ^{\dagger }L{\rm f}, &  & {\cal L}_{4}^{2}=i\delta _{2}{\rm \bar{f}%
}\gamma ^{\mu }U^{\dagger }\left( D_{\mu }U\right) R{\rm f}, \\ 
{\cal L}_{4}^{3}=i\delta _{3}{\rm \bar{f}}\gamma ^{\mu }\left( D_{\mu
}U\right) \tau ^{3}U^{\dagger }L{\rm f}+h.c., &  & {\cal L}_{4}^{5}=i\delta
_{5}{\rm \bar{f}}\gamma ^{\mu }\tau ^{3}U^{\dagger }\left( D_{\mu }U\right) R%
{\rm f}+h.c., \\ 
{\cal L}_{4}^{4}=i\delta _{4}{\rm \bar{f}}\gamma ^{\mu }U\tau ^{3}U^{\dagger
}\left( D_{\mu }U\right) \tau ^{3}U^{\dagger }L{\rm f}, &  & {\cal L}%
_{4}^{6}=i\delta _{6}{\rm \bar{f}}\gamma ^{\mu }\tau ^{3}U^{\dagger }\left(
D_{\mu }U\right) \tau ^{3}R{\rm f}, \\ 
{\cal L}_{4}^{7}=i\delta _{7}{\rm \bar{f}}\gamma ^{\mu }U\tau ^{3}U^{\dagger
}D_{\mu }^{L}L{\rm f}+h.c., &  & 
\end{array}
\]
where $L=\frac{1-\gamma ^{5}}{2},$ $R=\frac{1+\gamma ^{5}}{2}$ are the left
and right projectors. The matrix-valued field $U(x)$ is an $SU(2)$ matrix
containing the three Goldstone bosons associated to the spontaneous breaking
of the symmetry. The covariant derivatives appearing in the above operators
are 
\begin{eqnarray*}
D_{\mu }U &=&\partial _{\mu }U+ig\frac{{\bf \tau }}{2}{\bf \cdot W}_{\mu
}U-ig^{\prime }U\frac{\tau ^{3}}{2}B_{\mu }, \\
D_{\mu }^{L}{\rm f} &=&\left( \partial _{\mu }+ig\frac{{\bf \tau }}{2}{\bf %
\cdot W}_{\mu }+ig^{\prime }\frac{1}{6}B_{\mu }+ig_{s}\frac{{\bf \lambda }}{2%
}{\bf \cdot G}_{\mu }\right) {\rm f}, \\
D_{\mu }^{R}{\rm f} &=&\left( \partial _{\mu }+i\frac{g^{\prime }}{2}\left(
\tau ^{3}+\frac{1}{3}\right) B_{\mu }+ig_{s}\frac{{\bf \lambda }}{2}{\bf %
\cdot G}_{\mu }\right) {\rm f}.
\end{eqnarray*}
Finally, ${\rm f}$ is a weak doublet of matter fields ($(t,b)$ in our case).
Generation mixing has been neglected. The above operators contribute to the
different gauge boson-fermion-fermion vertices as indicated in table 1.

\begin{table}[tbp]
\centering
\begin{tabular}{|c|c|}
\hline
Vertex & Feynman Rule \\ \hline
$\bar{t}gt$ & $-ig_{s}\frac{\lambda }{2}^{a}\gamma _{\mu }\left( 1+2\delta
_{7}L\right) $ \\ \hline
$\bar{b}gb$ & $-ig_{s}\frac{\lambda }{2}^{a}\gamma _{\mu }\left( 1-2\delta
_{7}L\right) $ \\ \hline
$\bar{t}W^{+}b$ & $-\frac{i}{\sqrt{2}}g\gamma _{\mu }\left(
g_{L}L+g_{R}R\right) $ \\ \hline
$\bar{b}W^{-}t$ & $-\frac{i}{\sqrt{2}}g\gamma _{\mu }\left( g_{L}^{\ast
}L+g_{R}^{\ast }R\right) $ \\ \hline
\end{tabular}
\caption{Feynman rules for the vertices appearing in the subprocesses of
Figs.(\ref{t+b+}) and (\ref{t-b-}).}
\label{vertices}
\end{table}

In addition, the operator ${\cal L}_{4}^{7}$ also contributes to the quark
self energies and to the counterterms required to guarantee the on-shell
renormalization conditions \cite{Bagan}, but when we take into account all
these contributions, $\delta _{7}$ vanish from the observables in the
present case. It should be noted, however, that the internal quark line in
the diagrams in Figs.(\ref{t+b+}) and (\ref{t-b-}) are never on-shell and
the use of the equations of motion to eliminate ${\cal L}_{4}^{7}$, is a
priori not justified. The net effect of the electroweak effective lagrangian
in the charged current sector can thus be summarized, to the order we have
considered, in the effective couplings $g_{L}$ and $g_{R}$. 
\begin{equation}
g_{L}=1+\delta g_{L}=1-(\delta _{1}+\delta _{4})\qquad g_{R}=\delta
g_{R}=\delta _{2}-\delta _{6}.
\end{equation}

At present not much is known from direct measurements for the $t\rightarrow
b $ effective coupling. This is perhaps best evidenced by the fact that the
current experimental results for the (left-handed) $V_{tb}$ matrix element
give \cite{PDG} 
\begin{equation}
\frac{|V_{tb}|^{2}}{|V_{td}|^{2}+|V_{ts}|^{2}+|V_{tb}|^{2}}%
=0.94_{-0.24}^{+0.31}.
\end{equation}
It should be emphasized that these are the `measured' or `effective' values
of the CKM matrix elements, and that they do not necessarily correspond,
even in the Standard Model, to the entries of a unitary matrix on account of
the presence of radiative corrections, even though these deviations with
respect to unitary are expected to be small unless new physics is present.
At the Tevatron, it is said, the left-handed couplings are expected to be
eventually measured with a 5\% accuracy \cite{TEVA}, but as mentioned above
this depends on absolute scale normalizations which are hard to pin down,
even after a full two-loop QCD calculation.

As far as experimental bounds for the right handed effective couplings is
concerned no direct relevant bounds exist, the more stringent ones are
indirect and come from the measurements on the $b\rightarrow s\gamma $ decay
at CLEO \cite{CLEO}. Due to a $m_{t}/m_{b}$ enhancement of the chirality
flipping contribution, a particular combination of mixing angles and $\kappa
_{R}^{CC}$ can be found. The authors of \cite{LPY} reach the conclusion that 
$|{\rm Re}(\kappa _{R}^{CC})|\leq 0.4\times 10^{-2}$. However, considering $%
\kappa _{R}^{CC}$ as a matrix in generation space, this bound only
constraints the $tb$ element. Other effective couplings involving the top
remain virtually unrestricted from the data. Nonetheless the previous bound
on the right-handed coupling is a very stringent one. It is fairly clear
that an hadron machine such as the LHC will never be able to compete with
such a precision. Yet, the measurement would be a direct one, not through
loop corrections. Equally important is that it will yield information on the 
$ts$ and $td$ elements too, by just replacing the quark exchanged in the $t$%
-channel in Fig.\ref{t+b+} (b).

\section{The effective-$W$ approximation}

The calculations presented in this work are carried out in the framework of
the so-called effective-$W$ approximation that is the translation to the
present case of the familiar Weisz\"{a}cker-Williams\cite{WW} approximation
for photons.

Known to be accurate at high energies (see e.g. \cite{FMNR} for a discussion
on accuracy and improvements) and very convenient, this approach is
computationally simple and has all the attractive physical interpretation of
the parton model. One certainly would expect that the translation of the
Weisz\"{a}cker-Williams approximation to $W$'s is s good one at sufficiently
high energies. In LHC physics it has been amply used in the context of $W$$W$%
, $W$$Z$ or $W$$\gamma $ scattering. (See e.g. \cite{madrid} for a very
recent application and references.) In \cite{EffWapp}, the author claims
without much elaboration that the approximation works well for colliders at
20 TeV or higher energy. LHC falls somewhat short of this energy, so the
applicability becomes a bit problematic. This has motivated us to look into
this question in somewhat more detail. We shall later discuss to what extent
the above expectations are fulfilled.

In this work the production of polarized tops is considered. As we shall see
in the appendix, within the frame of the effective-$W$ approximation this is
absolutely necessary if one wishes to set bounds on the right-handed
effective couplings. In doing so we have found results which are somewhat at
variance with the recent work reported in \cite{parke}. A priori it would
not be clear whether the difference can be blamed on the effective-$W$
approximation itself or in the use of different kinematical cuts (which in
turn are somewhat forced upon us by the same effective-$W$ approximation).
Because of this, we shall review in the coming pages not just total cross
sections but cross sections for the production of polarized tops and quite
detailed $p_{T}$ and angular distributions and compare the results obtained
using the effective-$W$ approximation to the ones of the exact calculation 
\cite{exact} we performed recently.

In order to calculate the cross section of the process $pp\rightarrow t\bar{b%
}$ we have used the CTEQ4 structure functions \cite{CTEQ4} to determine the
probability of extracting a parton with a given fraction of momenta from the
proton. The $u$ and $d$-type partons then radiate a $W^-$ or $W^+$ boson,
respectively.

In the effective-$W$ approximation these $W$ bosons (both longitudinal and
transverse) are treated as partons from the proton, carrying a fraction of
the quark momentum and thus of the momentum of the proton. The $W$ parton
distribution function is, roughly speaking, the probability of producing a $%
W $ with such a fraction of the momentum. In the spirit of the
Weisz\"{a}ker-Williams approximation, the cross-section for the process $%
pp\rightarrow t\bar{b}X$ (for instance) is approximated by the product of
cross sections 
\begin{equation}
\int_{0}^{1}dy\int_{0}^{1}d\hat{x}\int d^{2}k_{T}d^{2}k_{T}^{\prime }\sigma
\left( pp\rightarrow W\left( k\right) g\left( k^{\prime }\right) \right)
\left( \frac{1}{k^{\prime 2}}\right) ^{2}\left( \frac{1}{k^{2}-M_{W}^{2}}%
\right) ^{2}\hat{\sigma}\left( k,k^{\prime }\right) ,  \label{weisw}
\end{equation}
where $\hat{\sigma}$ is the physical cross-section for the subprocess; $%
Wg\rightarrow t\bar{b}$ in our case. In this subprocess cross section, both
the $W$ and the gluon are assumed to be on-shell, i.e. we have a physical,
gauge independent, cross section. Of course the $W$ is never on-shell.
Kinematically, the $W$ has a space-like four momentum, and it is off its
mass shell by an amount which is, at least, $M_{W}^{2}$. However, at the
energies which are characteristic of the LHC, one expects the error to be
small. The variables $\hat{x}$ and $y$ are the fractions of the proton
energy carried by the $W$ and gluon, respectively. $k_{T}$ and $%
k_{T}^{\prime }$ are the respective transverse momenta of $W$ and gluon.
Hereafter when we talk about components of a given four-momentum we consider
them given in the LAB frame (center-of-mass frame of the two protons). The $%
W $ momentum can be written as $k=(\hat{x}E,k_{T},\omega )$, where $\omega $
is the longitudinal momentum of the $W$, $E$ is the energy of the $u$
parton, and $\hat{x}$ the fraction of the parton energy carried by the $W$. $%
E$ is related to the total energy of the proton by $E=xE_{P}$, $E_{P}$ being
the proton energy in the LAB frame.

Energy-momentum conservation in the vertex requires that, if the emerging
`spectator' quark (a $d$ quark if the parton radiating the $W$ is a $u$
quark) is to be on-shell, the squared four-momentum of the $W$ is negative
or zero. In this case, $\omega =E-\sqrt{(1-\hat{x})^{2}E^{2}-k_{T}^{2}}$,
and the cut for the integration over $k_{T}$ is, of course, $(1-\hat{x})E$.
The $W$ is off-shell by at least an amount $M_{W}^{2}$. On the other hand,
one may decide to set the $W$ on-shell, since the subprocess cross-section
is after all computed for on-shell $W$'s (and only in this case it is
physically meaningful and gauge independent). In this case, the longitudinal
momentum carried by the $W$ is $\omega =\sqrt{\hat{x}%
^{2}E^{2}-M_{W}^{2}-k_{T}^{2}}$, and the cut on the integration over
transverse momenta will be $\sqrt{\hat{x}^{2}E^{2}-M_{W}^{2}}$. This latter
choice sets the `spectator' quark off-shell by a virtuality of order $%
M_{W}^{2}$ (recall that previously it was the $W$ itself which was
off-shell).

In either case, we rewrite Eq.(\ref{weisw}) in the form 
\begin{equation}
\int_{0}^{1}dyf_{g}(y)\int_{0}^{1}dxf_{u}(x)\int_{0}^{1}d\hat{x}f_{W}(\hat{x}%
,E)\hat{\sigma}(\hat{x},y),  \label{pdfs}
\end{equation}
with $f_{g}$ and $f_{u}$ the parton distribution functions of the gluon and $%
u$ type parton. Equations (\ref{weisw}) and (\ref{pdfs}) define the $W$
parton distribution function $f_{W}$.

In Eq.(\ref{pdfs}) we have replaced the $W$ and gluon momenta, $k$ and $%
k^{\prime }$, by their $z$ components. The approximation thus involves
neglecting the transverse momenta in $\hat{\sigma}$, which is integrated
over. Depending on whether one chooses to take $k^{2}=0$ or $k^{2}=M_{W}^{2}$
for the $W$, this amounts to replacing the intermediate boson four momenta
by $(\hat{x}E,0,0,\hat{x}E)$ or $(\hat{x}E,0,0,\sqrt{\hat{x}%
^{2}E^{2}-M_{W}^{2}})$, respectively. The integration over the transverse
momenta is represented by the $W$ parton distribution functions $f_{W}$.
This will of course be a good approximation inasmuch as the process is
strongly dominated by $k_{T}=0$.

In passing from Eq.(\ref{weisw}) to Eq.(\ref{pdfs}) one averages over the
possible values of the transverse momenta. For `normal' partons (the gluon,
for instance) this leads to a mass singularity as $k_{T}\rightarrow 0$; the
distribution is clearly peaked at low values of $k_{T}$, leading to the
familiar logarithmic dependence on the scale. On the other hand, for the $W$
the mass singularity is absent due to the mass in the propagator. There is
thus a natural spread in the distribution of $k_{T}$ which makes the
effective-$W$ approximation less accurate. Obviously the approximation
becomes better the larger the value of $E$ is. As we mentioned, Dawson\cite
{EffWapp} and others \cite{KRR,Johnson} somehow estimated the accuracy of
the approximation. Half the cross section for transverse $W$'s comes from
angles $\theta \leq \sqrt{M_{W}/2E}$, and the cross section is even more
collimated for longitudinal $W$'s. We have set a cut in the sub-process
invariant mass to guarantee the validity of the effective-$W$ approximation.
If we neglect altogether the $W$ mass, the invariant mass of the $Wg$ system
(or, for that matter, the top-anti-bottom system) is $4\hat{x}yEE_{P}$
(since the $W$ and the gluon have opposite directions within this
approximation). A lower cut in the invariant mass is a cut in $\hat{x}$, $y$
and, in particular, on $E$. We shall in the following present results for a
couple of cuts in the invariant mass, 500 GeV and 100 GeV. The validity of
the effective-$W$ approximation appears questionable in the second case, not
so much in the first one. However one of our conclusions will be that no
matter the cut that one sets in the invariant mass there are problems. The
limiting factor will in fact be the LHC energy.

The upper limit for the integral over $k_{T}$ sets the scale normalizing the
(logarithmic) dependence on $M_{W}$ of the structure functions. It is
somewhat ambiguous to set a given value for this scale. Some authors (see
e.g. \cite{madrid,mele} take $k_{T}^{max}=E^{2}$ (the energy of the $u$ or
the gluon in its center-of-mass frame) while others take $4E^{2}$ \cite
{EffWapp}. It should be borne in mind that the uncertainty associated to
using one value or another, while nominally subdominant is again not so
small at LHC energies, so the difference matters to some extent. The
relevant expressions for $f_{W}$ that we have used can be found in \cite
{EffWapp}. Next to leading calculations exist in the literature \cite
{Johnson}.

There are, in fact, two different parton distribution functions for the $W$,
one for transverse and another one for longitudinal vector bosons, $%
f_{W_{T}} $ and $f_{W_{L}}$ respectively. Needless to say that the
distinction between transverse and longitudinal $W$'s is not Lorentz
invariant--- a transverse photon may turn into a combination of transverse
and longitudinal after a Lorentz transformation. However, provided that the
changes of reference frame involve only boosts in the $z$ direction, the
transverse degrees of freedom remain transverse, while the longitudinal ones
mix with the temporal ones, but gauge invariance of the physical amplitude
for the sub-process does guarantee that the correct result is preserved.
Since in the effective-$W$ approximation all the dynamics in the initial
state takes place in the $z$ direction one needs not worry in which precise
reference frame these distribution functions are defined.

It turns out to be absolutely crucial for our purposes to keep the parton
distribution functions for the $W$ as given, for instance in \cite{EffWapp},
without attempting to approximate them by assuming that $E>>M_{W}$. For
instance, one often finds in the literature the following approximate
expression for $f_{W_{T}}$ 
\begin{equation}
f_{W_{T}}(\hat{x})\simeq \frac{g^{2}}{\left( 4\pi \right) ^{2}}\frac{\hat{x}%
^{2}+2\left( 1-\hat{x}\right) }{2\hat{x}}\log \left( \frac{4E^{2}}{M_{W}^{2}}%
\right) ,
\end{equation}
Using this expression instead of the one given in \cite{EffWapp}
overestimates the total cross section by at least a factor five. To reach
this conclusion we compare our results to our exact analytical calculation
presented in \cite{exact}. When one looks in detail the kinematical regions
that matter, the energy of the LHC is just not large enough to grant the
approximation, and increasing the cut of 500 GeV in the sub-process
invariant mass further does not really help. It is thus essential to use an
expression for $f_{W_{T}}(\hat{x})$ which is valid over all ranges of $\hat{x%
}$. Even in this case the results are not fully satisfactory as we shall see
in a moment.

Regarding the longitudinal $W$ parton distribution function $f_{W_{L}}(\hat{x%
})$ we have realized that the complete expression given in \cite{EffWapp} is
incorrect because we have found numerically that it is not positive definite
as it should. Despite of that we have found that the approximate expression
(which is evidently positive definite) given in the same work 
\begin{equation}
f_{W_{L}}(\hat{x})\simeq \frac{g^{2}}{\left( 4\pi \right) ^{2}}\frac{1-\hat{x%
}}{\hat{x}},
\end{equation}
gives rise to sensible results when compared to the ones obtained in the
exact calculation. Because of that we have proceeded to used it, obtaining
that the corresponding contribution to the final result is much smaller that
the one coming from the transverse sector (about a 10 \% in our case).

The fact that the longitudinal degrees of freedom should be subdominant can
be understood on the following grounds. At large energies one can
approximate the polarization vector $\varepsilon _{L}^{\mu }$ by $k^{\mu
}/M_{W}$. Taking into account that the longitudinal vector bosons couple to
the light quarks, use of the equations of motion forces this term to vanish
or be negligible. In fact, the only term that may survive is the one that is
subdominant when one uses the above approximation for $\varepsilon _{L}^{\mu
}$, namely 
\begin{equation}
\varepsilon _{L}^{\mu }=\frac{k^{\mu }}{M_{W}}+\frac{\left| {\bf k}\right|
-k^{0}}{M_{W}}\left( 1,0,0,-1\right) ,
\end{equation}
We have used $\varepsilon _{L}^{\mu }=\frac{\left| {\bf k}\right| -k^{0}}{%
M_{W}}\left( 1,0,0,-1\right) $ in all cases.

At this point one must commit oneself to a given choice for the $k^{2}$ of
the virtual $W$ as it is impossible to keep both the two light quarks and
the $W$ on shell. We have found, in fact, that for the final results it
hardly matters whether one uses $k^{2}=0$ or $k^{2}=M_{W}^2$ (or,
presumably, anything in between). Here we shall present results of the
latter option and we will postpone a mode detailed discussion on this issue
to another paper. The main argument in favour of this choice is that, except
for the fact that one of the external legs is off-shell by an amount which
is nevertheless small compared to the relevant energies, it is the one that
matches smoothly the formulae for the $W$ parton distribution functions
given in \cite{EffWapp}. In that case $k^{0}$ is always bigger than $M_{W}$
and factors appearing in the $W$ parton distribution functions of the form $%
\left( k^{0}-M_{W}\right) ^{-1/2}$ are well defined real numbers. If we want
to take $k^{2}=0$ and use at the same time the results of \cite{EffWapp} we
have to impose an artificial cutoff enforcing $k^{0}>M_{W}$ in order to
assure a sensible cross section

The other obvious approximation involved in using the effective-$W$
approximation is the neglection of the crossed interference term between
longitudinal and transverse $W$'s. In the case at hand, the cross sections
of the elementary subprocesses of Fig.\ref{t+b+} are presented in the
Appendix and it is not difficult to check that they are of the same order.
However, the arguments we have given previously concerning the dominance of
the transverse $W$ make the longitudinal contribution subdominant. However,
one should still worry about the interference term, which could easily give
a correction of the order of 30 \%. Fortunately it can be seen that
integration over the azimutal angle makes the interference term to
approximately vanish\cite{kauf,Johnson}. So in fact, the neglection of the
interference term is not a bad approximation at all.

\section{Cross Section for Single Top Production}

\label{cross}We have thus proceeded as follows. We have multiplied the
parton distribution function of a gluon of a given momenta from the first
proton by the sum of parton distribution functions for obtaining a $u$ type
quark from the second proton. Then we have multiplied this result by the
probability of obtaining an on-shell transversal $W^{+}$ from those partons.
We have repeated the process for a longitudinal vector boson. These results
are then multiplied by the cross sections of the subprocesses of Fig.\ref
{t+b+} corresponding to transversal or longitudinal $W^{+}$, respectively.
At the end, these two partial results are add up to obtain the total $%
pp\rightarrow t\bar{b}$ cross section.

Since typically, the top quark decays weakly well before strong interactions
become relevant, we can in principle measure its polarization state with
virtually no contamination of strong interactions (see e.g. \cite{parke} for
discussions on how this could be done). For this reason we have considered
polarized cross sections and provide general formulas for the production of
polarized tops in a general spin frame (within the context and limitations
of the effective-$W$ approximation). The mass of the $b$ quark has been
maintained all the way.

To calculate the event production rate corresponding to different
observables and compare them with the theoretical predictions we have used
the integrating montecarlo program VEGAS \cite{vegas}. We present results
after one year run at full luminosity in one detector (100 ${\rm fb}^{-1}$
at LHC). The scale used for $\alpha _{s}$ is the invariant mass of the
partons (the gluon and the light quark). For the purposes of this work a
more detailed study on this scale dependence does not seem necessary.

Let us start by discussing the experimental cuts. Due to geometrical
detector constraints we adopt a pseudorapidity cut $|\eta |<2.5$ both for
the top and bottom. This corresponds to approximately 10 degrees from the $z$
axis. As for $p_{T}$ we have taken the cut $|p_{T}|>$ 30 GeV. Within the
effective-$W$ approximation the $W$ and gluon transverse momenta are
neglected; this implies that the top and bottom $p_{T}$ are identical. This
last assumption is not valid in an exact calculation and to what extent this
changes the results depends on the cuts selected. We have also implemented
an angular isolation cut for the top and anti-bottom (or anti-top and
bottom) of 20 degrees. These cuts are relatively mild, they reduce the cross
section by about a factor three.

In single top production a distinction is often made between $2\rightarrow 2$
and $2\rightarrow 3$ processes. The latter corresponds, in fact, to the
process we have been discussing, the one represented in Fig.\ref{t+b+}, in
which a gluon from the sea splits into a $b$ $\bar{b}$ pair. In the $%
2\rightarrow 2$ process the $b$ quark is assumed to be extracted from the
sea of the proton. Of course the distinction between the two processes is
merely kinematical and somewhat arbitrary. In the remains of the proton a $%
\bar{b}$ must be present, given that the proton has no net $b$ content and
thus the final state is also identical to the one we have been discussing.
The values of the total cross sections presented in the introduction
correspond to the kinematical cuts used in \cite{SSW}. In the framework of
our approximation all partons are deemed to have zero transverse momentum
and hence the detection of a $\bar{b}$ in the fiducial zone, above the
angular and/or $p_{T}$ cuts, necessarily indicates that the $\bar{b}$
originates from the `hard' sub-process. In \cite{exact} we discuss this
issue in more detail and a justification of the above signal for single top
production is given. Let us just mention here that if we were to use the
signal suggested by Willenbrock and coworkers (only one bottom tag in the
final state) the effective-$W$ approximation would be completely useless.

As for the cut in the invariant mass we have alluded to before, we have used
two values, namely 500 GeV and 100 GeV. Both are somewhat extreme. The first
one eliminates many events (the total cross section is reduced by an order
of magnitude compared to the other cut), while the latter renders the
effective-$W$ approximation even more questionable. As we will see the
comparison with the exact calculation does not really improve when the cut
on the invariant mass is raised.

\begin{figure}[tbp]
\epsfysize=7.cm %\epsfxsize=7.0cm
\centerline{\epsffile{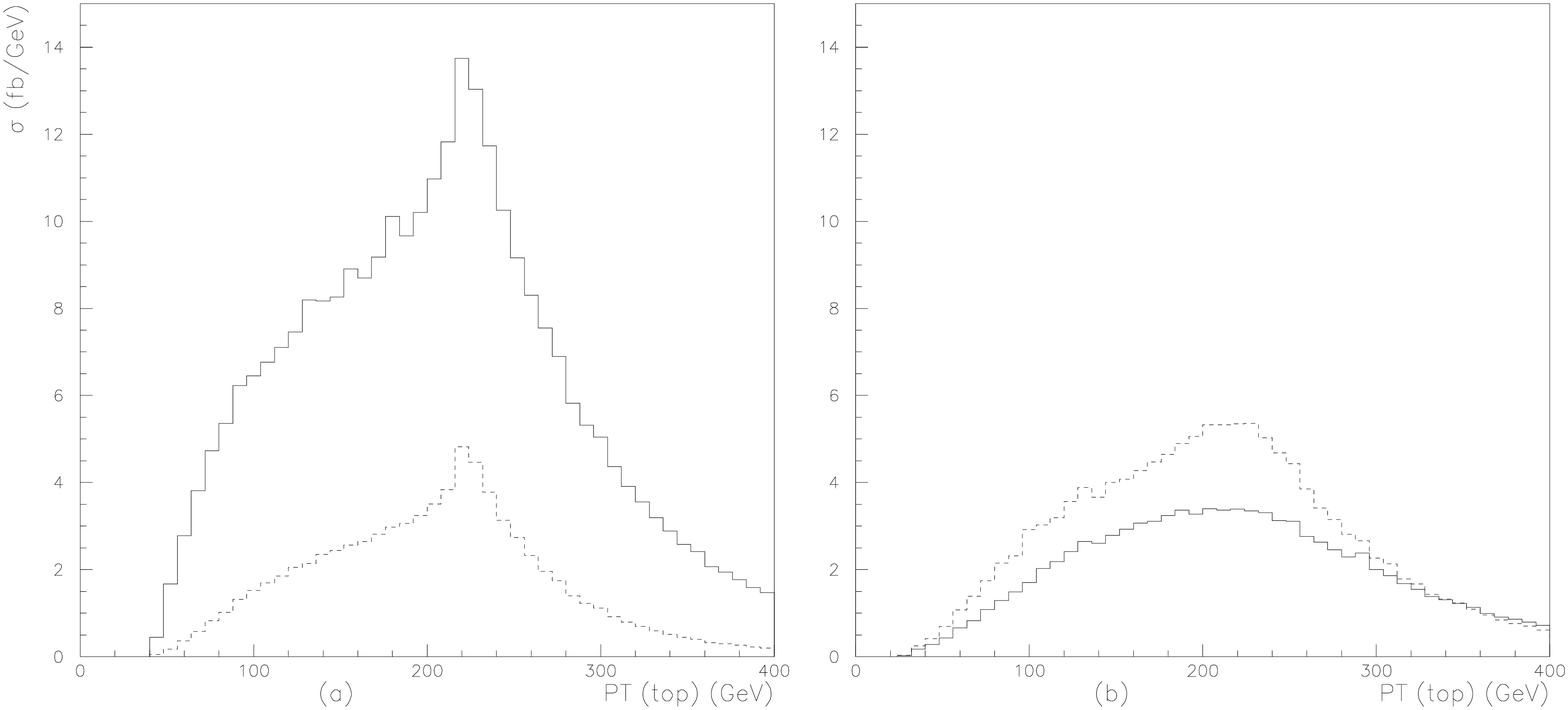}}
\caption{Differential cross section of (single) tops produced at the LHC vs.
transversal momentum in the Standard Model. The solid (dotted) line
corresponds to left (right) polarized top production. The subprocesses
contributing to these histograms have been calculated at tree level in the
electroweak theory with a cut of 500 Gev. in the top anti-bottom invariant
mass. In the figure we show the results of the calculations for polarized
top production in the LAB helicity basis. These predictions (a) are compared
to those of the exact calculation (b). We see that the effective-W
approximation fails to produce the correct polarization behaviour}
\label{pttopeffwexact500_lab_lr_gl=1_gr=0}
\end{figure}

\begin{figure}[tbp]
\epsfysize=7.cm %\epsfxsize=7.0cm
\centerline{\epsffile{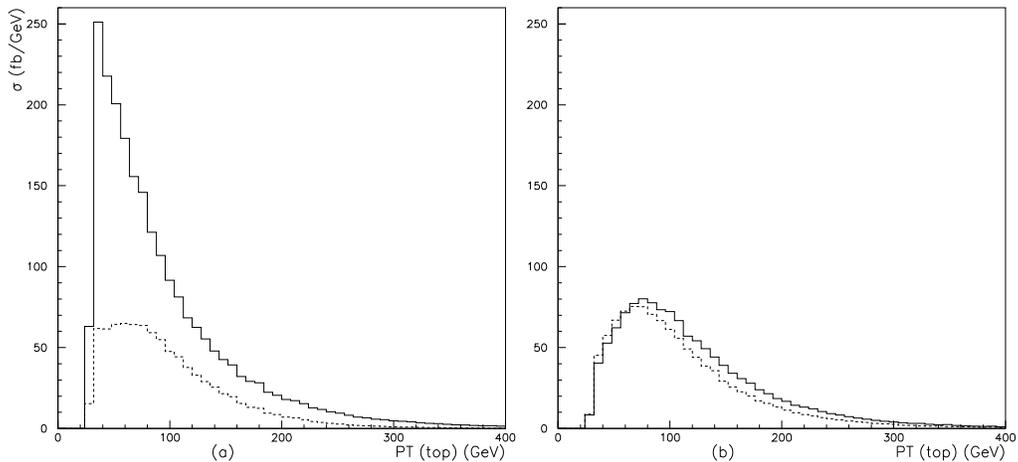}}
\caption{The same plots as in Fig.\ref{pttopeffwexact500_lab_lr_gl=1_gr=0}
but with a cut of 100 GeV in the invariant mass of the top anti-bottom
system. Again the effective-$W$ approximation fails to account for the
correct $P_{T}$ distributions.}
\label{pttopeffwexact100_lab_lr_gl=1_gr=0}
\end{figure}

We shall start by considering the Standard Model tree-level predictions
concerning single top production. In Table \ref{results} and Figs. \ref
{pttopeffwexact500_lab_lr_gl=1_gr=0} and \ref
{pttopeffwexact100_lab_lr_gl=1_gr=0} we present our numerical results for
production of polarized tops in the LAB helicity basis and compare them to
an the exact calculation of the cross-section for the two cuts on the
invariant mass. From these two figures we can see that the results obtained
using the effective-$W$ approximation differ significantly from the ones
obtained from an exact calculation. In particular we immediately observe
that the effective-$W$ approximation systematically leads to many more
left-handed tops that right-handed ones while the situation in an exact
calculation is more complicated (we of course mean negative and positive
helicity when we talk of left or right tops, chirality simply does not make
sense for such a massive particle).

Indeed, the percentage of left tops with respect to the total production
depends, in an exact calculation, critically on the cuts imposed. We have
observed numerically in the exact calculation that left tops are produced in
excess to right tops for low invariant mass and viceversa for high invariant
mass. This is seems counterintuitive since we expect that at high invariant
masses we could approximate the top as massless and therefore we expect it
to be produced mainly left-handed polarized. However we have observed that
in the particular mechanism considered for single top production in the LHC
the average top energy is not much higher than its mass for such argument to
hold. It is important to remark also that in order to compare the effective-$%
W$ approximation results to the ones of an exact calculation no cuts are
placed on the spectator quark in the last one (recall that the spectator
quark is invisible in the effective-$W$ approximation).

\begin{figure}[tbp]
\epsfysize=7.cm %\epsfxsize=7.0cm
\centerline{\epsffile{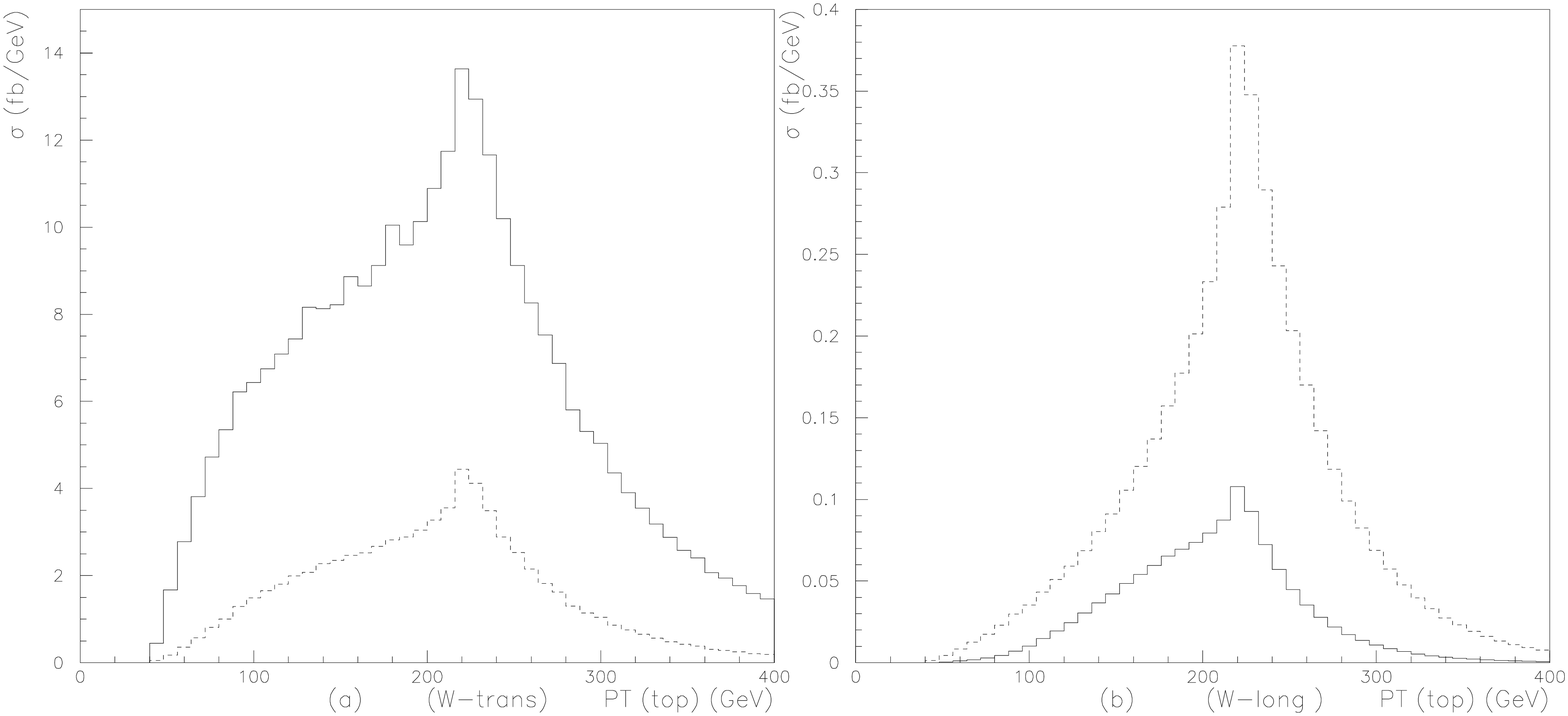}}
\caption{Differential cross section of (single) tops produced at the LHC vs.
transversal momentum in the Standard Model. The solid (dotted) line
corresponds to left (right) polarized top production. The subprocesses
contributing to these histograms have been calculated at tree level in the
electroweak theory with a cut of 500 GeV in the top anti-bottom invariant
mass. In the figure we show the results of the calculations for polarized
top production in the LAB helicity basis. Both figures are calculated in the
effective-W approximation with (a), (b) showing the contribution coming from
transverse and longitudinal polarized $W$ respectively. It is clear that
transverse $W$ largely dominates (notice the different vertical scale).}
\label{pttopeffwtranslong_lab_lr_gl=1_gr=0}
\end{figure}

\begin{figure}[tbp]
\epsfysize=7.cm %\epsfxsize=7.0cm
\centerline{\epsffile{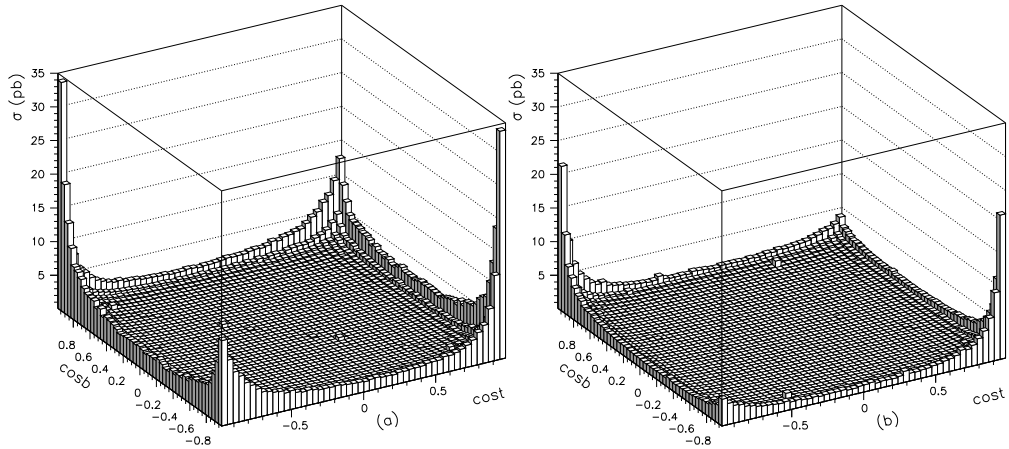}}
\caption{Expected angular distribution for single top production produced at
the LHC in the Standard Model calculated in the effective $W$ approximation
(a) and with an exact calculation (b) .}
\label{ctcbeffwexact_unpol_gl=1_gr=0}
\end{figure}

The fact that the effective-$W$ approximation produces many more left tops
than right ones can to some extent be understood on pure kinematical and
angular momentum conservation arguments. Take for instance a relatively
large cut in the invariant mass (such as 500 GeV). This automatically forces
a back to back kinematics in the emerging top and anti-bottom (see Fig.\ref
{ctcbeffwexact_unpol_gl=1_gr=0}). Since we know that the process is very
much dominated by transverse $W$ and the bulk of the processes take place in
the forward direction, angular momentum conservation forces the top to
emerge with a negative helicity. Indeed, the contribution of longitudinally
polarized $W$'s does favour right handed tops, but it is numerically
irrelevant (see Fig.\ref{pttopeffwtranslong_lab_lr_gl=1_gr=0}). The
situation still persists at lower values of the invariant mass, where both
back to back and same direction top-anti-bottom pairs are produced, but then
the exact calculation also does give marginally more negatively polarized
tops. Of course the above arguments do not really apply to the exact
calculation, where the kinematics is a lot more involved.

\begin{table}[tbp]
\centering
\begin{tabular}{|c|c|c|}
\hline
assumption & $N_{-}$ & $N_{+}$ \\ \hline
eff-W ($\sqrt{s}>$500 GeV) & $2.45\times 10^{5}$ & $0.63\times 10^{5}$ \\ 
\hline
exact ($\sqrt{s}>$500 GeV) & $0.82\times 10^{5}$ & $1.09\times 10^{5}$ \\ 
\hline
eff-W ($\sqrt{s}>$100 GeV) & $18.0\times 10^{5}$ & $7.08\times 10^{5}$ \\ 
\hline
exact ($\sqrt{s}>$100 GeV) & $9.92\times 10^{5}$ & $8.52\times 10^{5}$ \\ 
\hline
\end{tabular}
\caption{Total number of events in single top production in the LAB helicity
frame. We show a comparison between results obtained with and without the
effective-$W$ approximation in the tree level SM. Values calculated with $%
p_{T}>30$ GeV., and $10^{\circ }<\protect\theta <170^{\circ }$. We present
the comparison for invariant mass cuts of $500$ and $100$ GeV.}
\label{results}
\end{table}

\begin{figure}[tbp]
\epsfysize=7.cm %\epsfxsize=7.0cm
\centerline{\epsffile{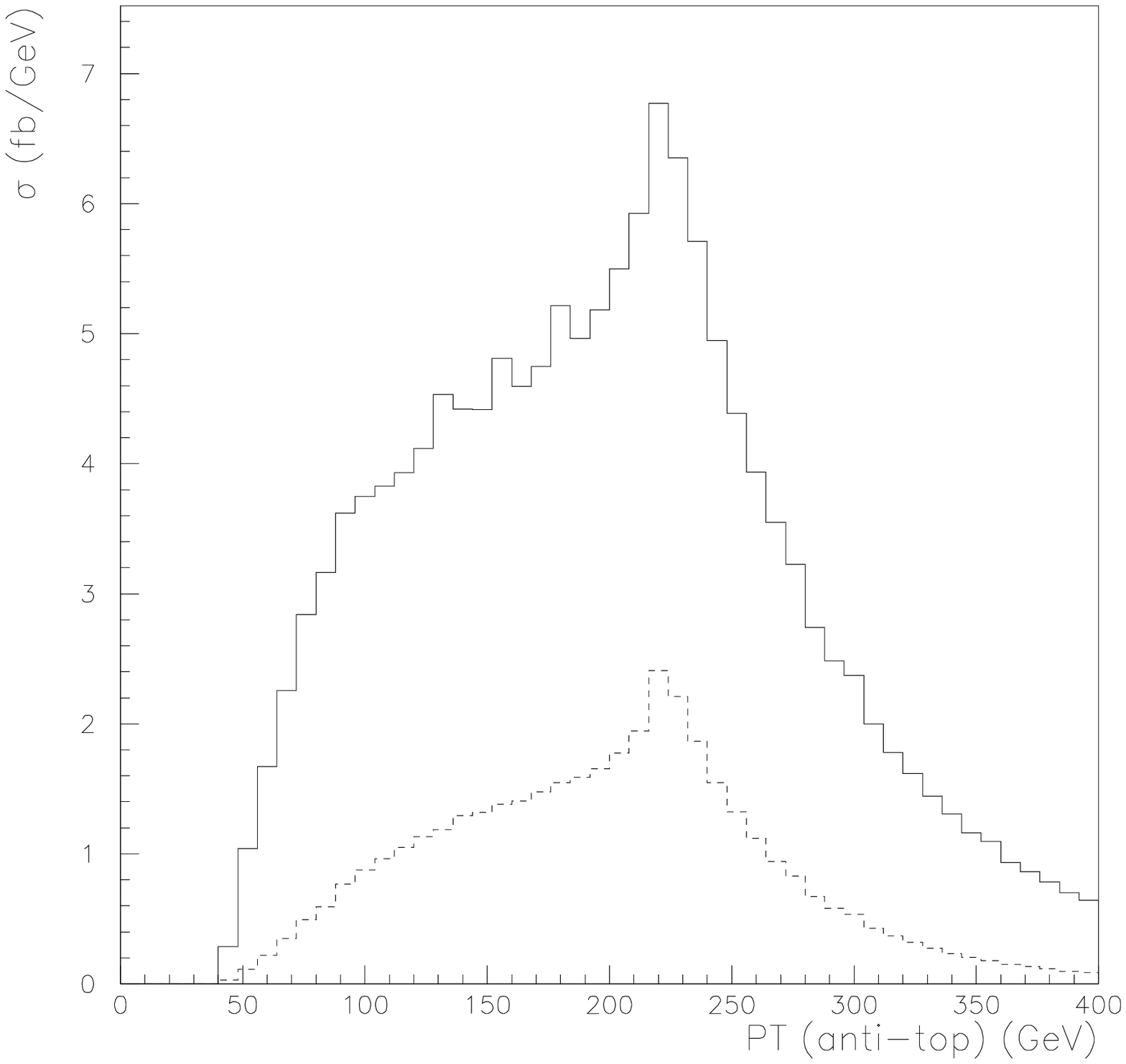}}
\caption{Differential cross section of (single) anti-top at the LHC vs.
transversal momentum at tree level in the Standard Model. The solid (dotted)
line corresponds to right (left) polarized anti-top production. Histograms
correspond to subprocesses calculated in the tree level electroweak
approximation in the LAB helicity frame within the effective-$W$
approximation approach.}
\label{ptatopeffw_lab_lr_gl=1_gr=0}
\end{figure}

We have also calculated single anti-top production. In Fig.\ref
{ptatopeffw_lab_lr_gl=1_gr=0} we show two different histograms corresponding
to the production of $\bar{t}$ with the two possible helicities in the LAB
frame and the 500 GeV cut on the invariant mass. All the histograms
correspond to the tree level electroweak approximation and clearly show that
single anti-top production is suppressed roughly by a factor of two with
respect to single top production. This feature is general and is due to the
different probability of extracting a $W^{-}$ from a proton as compared to
that of extracting a $W^{+}$. The relevant electroweak cross sections (see
Appendix) are symmetric under the interchange of particle by antiparticle
along with helicity flip.

In Fig. \ref{ctcbeffwexact_unpol_gl=1_gr=0} we plot the angular distribution
of single top production in the (tree-level) Standard Model and compare them
to the exact calculation \cite{exact} with an equivalent set of cuts. From
the inspection of this figure we see that as expected the distribution is
strongly peaked in the beam direction, with the probability of top and
anti-bottom being produced back to back bigger than produced parallel. The
exact calculation has similar features. The back to back production is
favoured with respect to the parallel one due to the 500 GeV. cut in the
invariant mass, as discussed. If we lower this cut in the exact calculation
(the effective-$W$ needs it) we observe that parallel configurations tend to
equal or exceed back-to-back ones (see \cite{exact})

\begin{figure}[tbp]
\epsfysize=7.cm %\epsfxsize=7.0cm
\centerline{\epsffile{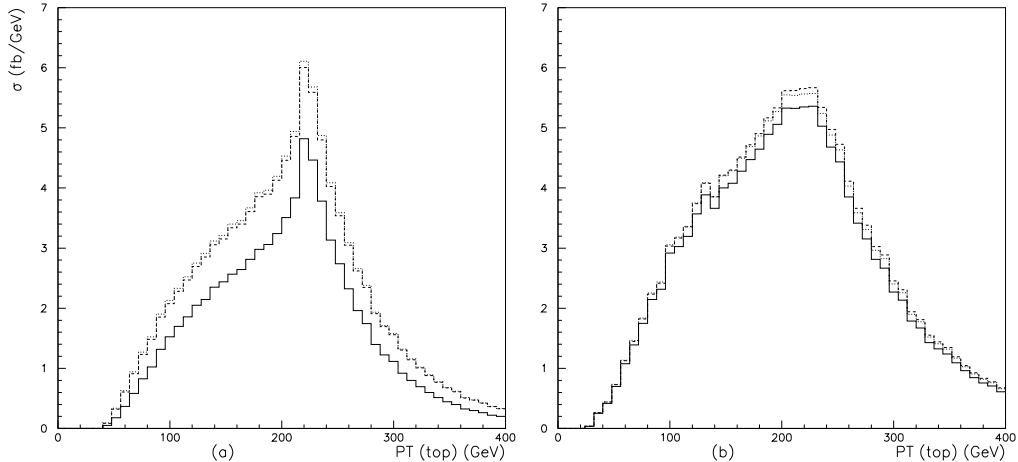}}
\caption{(a) Comparison between the tree Standard Model $p_{T}$ distribution
for single positively polarized top production (solid line) and the
corresponding obtained with a value $g_{R}=\pm 0.3$ for the effective right
handed coupling (dotted/dashed line) in the effective $W$ approximation. (b)
Same when the exact calculation is used.}
\label{pttopeffwexact_lab_r_gl=1_gr=0.3}
\end{figure}

As we see, the predictions from the effective-$W$ approximation coincides
only roughly with the exact calculation ones. The angular distribution is
similar, and the total cross section is in the same ballpark, but this is
about all. However, an intriguing feature of the approximation is that the
production of positively polarized (right) tops comes out in surprising good
agreement with the exact calculation. Indeed, let us now depart from the
tree-level Standard Model and consider non-zero values for $\delta g_{R}$.
In Fig. \ref{pttopeffwexact_lab_r_gl=1_gr=0.3} we present a comparison
between the effective-$W$ approximation and the exact calculation for
several values of the right effective couplings; the agreement is not bad.
The cut for the invariant mass is $500$ GeV. and the remaining cuts are the
ones used so far.

\section{Conclusions}

We have done a complete calculation of the subprocess cross sections for
polarized tops or anti-top production at the LHC including all mass
corrections and with general effective couplings $g_L$ and $g_R$. The
calculations presented are fully analytical.

Then we have used those results to analyzed the single top production
process at the LHC using the effective-$W$ approximation. That is,
considering that the W-boson is a real particle in order to calculate the
probability of obtaining a W-boson from a proton as a product of
probabilities. The effective-$W$ approximation has in its favour its
technical simplicity and a clear physical interpretation for those used to
the language of parton distribution function, being a generalization of the
Weisz\"{a}cker-Williams approximation for photons, which works well.

In this work we have shown that the effective-$W$ approximation is well
suited for angular distributions and works reasonably well (but not
exceedingly well) for total cross sections where it provides estimations
that are in the same ballpark than those provided by an exact calculation.
It fails however to reproduce the detailed $p_{T}$ distributions. It does
not reproduce well the fine details of the polarized top production,
systematically giving far too many left tops. In summary, it is not really
adequate for polarization studies and we urge the interested reader to use
the methods discussed in \cite{exact}.

\section{Acknowledgments}

It has been a pleasure to be able to present this work in the XXIX
International Meeting on Fundamental Physics, in a special session devoted
to the 60th birthday of Paco Yndur\'ain, to whom this work is dedicated,
with our best wishes of a long and productive career. We would like to thank
A.Dobado, M.J.Herrero, J.R.Pel\'aez and E.Ru\'\i z-Morales for multiple
discussions. J.M. acknowledges a fellowship from Generalitat de Catalunya,
grant 1998FI-00614. Financial support from grants AEN98-0431, 1998SGR 00026
and EURODAPHNE is greatly appreciated.

\appendix

\section{Effective Couplings and Mixed States}

Using our results (including the effective-$W$ approximation) we obtain that
the differential cross section matrix element at tree level can be written
as 
\begin{eqnarray}
\left| M\right| ^{2} &=&\left( \left| g_{R}\right| ^{2}+\left| g_{L}\right|
^{2}\right) a+\left( \left| g_{R}\right| ^{2}-\left| g_{L}\right|
^{2}\right) b_{n}+m_{b}m_{t}\frac{g_{R}^{\ast }g_{L}+g_{L}^{\ast }g_{R}}{2}c
\\
&=&\left( 
\begin{array}{cc}
g_{R}^{\ast } & g_{L}^{\ast }
\end{array}
\right) \left( 
\begin{array}{cc}
a+b_{n} & \frac{m_{b}m_{t}}{2}c \\ 
\frac{m_{b}m_{t}}{2}c & a-b_{n}
\end{array}
\right) \left( 
\begin{array}{c}
g_{R} \\ 
g_{L}
\end{array}
\right) \\
&\equiv &\left( 
\begin{array}{cc}
g_{R}^{\ast } & g_{L}^{\ast }
\end{array}
\right) A\left( 
\begin{array}{c}
g_{R} \\ 
g_{L}
\end{array}
\right) ,  \label{new}
\end{eqnarray}
where $a$, $b_{n}$, and $c$ are independent of the effective couplings $%
g_{R} $ and $g_{L}$ and $b_{n}$ is the only piece that depends on the top
spin four-vector $n.$ When $\left| M\right| ^{2}$ corresponds to the matrix
element of the subprocess the exact expressions for $a$, $b_{n}$, and $c$
can be obtained from the formulae given in the Appendix \ref{cs}. For the
matrix element of the whole process we have to multiply those expressions
with the corresponding $W$ and gluon parton distribution functions and make
the sums over parton species and polarizations but all this respects the
general form given in Eq.(\ref{new}). From Eq.(\ref{new}) we can observe
that $CP$ violating phases appear suppressed by the bottom mass but at the
same time they are enhanced by a $g_{L}$ factor. Hence, if one manages to
find a highly polarized spin basis there are some chances of having
experimentally observable effects. This is for sure not the case in the LAB
helicity basis as can be observed in Table \ref{results}. Also we see that
to determine $g_{R}$ we need to measure the top polarization.

Returning to the discussion of the general aspects of Eq.(\ref{new}) we
observe that $A$ is a symmetric matrix and then it is diagonalizable.
Moreover, from the positivity of $\left| M\right| ^{2}$ we immediately
arrive at the constraints 
\begin{eqnarray}
\det A &=&a^{2}-b_{n}^{2}-\frac{m_{b}^{2}m_{t}^{2}c^{2}}{4}\geq 0,
\label{constr1} \\
\frac{1}{2}TrA &=&a\geq 0.
\end{eqnarray}
In order to have a 100\% polarized top we need a spin four-vector $n$ that
saturates constrain (\ref{constr1}) for each kinematical situation, that is
we need $A$ to have a zero eigenvalue. In general such $n$ need not exist
and, should it exist, is in any case independent of the anomalous couplings $%
g_{R}$ and $g_{L}$ (this will not hold if we give up the effective-$W$
approximation). Moreover, provided this $n$ exists there is only one
solution (up to a global complex normalization factor) for the pair $\left(
g_{R},g_{L}\right) $ to the equation $\left| M\right| ^{2}=0,$ or
equivalently to the eigenvector equation 
\begin{equation}
A\left( 
\begin{array}{c}
g_{R} \\ 
g_{L}
\end{array}
\right) =0,  \label{100}
\end{equation}
To illustrate these considerations let us give an example: In the unphysical
situation where $m_{t}\rightarrow 0$ it can be shown that there exists two
solutions to the saturated constraint (\ref{constr1}), namely 
\begin{equation}
m_{t}n^{\mu }\rightarrow \pm \left( \left| \vec{p}_{1}\right| ,p_{1}^{0}%
\frac{\vec{p}_{1}}{\left| \vec{p}_{1}\right| }\right) ,
\end{equation}
once we have found this result we plug it in the expression (\ref{100}) and
we find the solutions $\left( 0,g_{L}\right) $ with $g_{L}$ arbitrary for
the + sign and $\left( g_{R},0\right) $ with $g_{R}$ arbitrary for the -
sign. That is, physically we have zero probability of producing a right
handed top when we have only a left handed coupling and viceversa when we
have only a right handed coupling. Note that if our theory had a massless
top and whatever non-null anomalous couplings $g_{R}$ and $g_{L}$ then there
would be no direction of 100\% polarization. This can be understood
remembering that the top particle forms in general an entangled state with
the other particles of the process and since we are tracing over the unknown
spin degrees of freedom of those particles we do not expect in general to
end up with a top in a pure polarized state, although this is not impossible
as it is shown the in above example.

\section{Subprocesses cross sections}

\label{cs}In order to write the cross section of the subprocess, we define
the spin four-vector corresponding to the spin in the $\hat{n}$ direction as 
\[
n^{\mu }\equiv \frac{1}{\sqrt{\left( p_{1}^{0}\right) ^{2}-\left( \vec{p}%
_{1}\cdot \hat{n}\right) ^{2}}}\left( \vec{p}_{1}\cdot \hat{n},p_{1}^{0}\hat{%
n}\right) , 
\]
with the properties 
\begin{eqnarray*}
n^{2} &=&-1, \\
\left( n\cdot p_{1}\right) &=&0,
\end{eqnarray*}
which reduces in the case of $\pm $helicity ($\hat{n}=\pm \frac{\vec{p}_{1}}{%
\left| \vec{p}_{1}\right| }$) to 
\[
n^{\mu }\equiv \pm \frac{1}{m_{t}}\left( \left| \vec{p}_{1}\right| ,p_{1}^{0}%
\frac{\vec{p}_{1}}{\left| \vec{p}_{1}\right| }\right) , 
\]
we have the differential cross section of the subprocess for single top
production 
\begin{eqnarray*}
d\sigma &=&f_{g}(y)f_{u}(x)f_{W}(\hat{x},E)dxdyd\hat{x}\delta ^{4}\left(
k_{1}+k_{2}-p_{1}-p_{2}\right) \\
&&\times \frac{1}{4\left| k_{2}^{0}\vec{k}_{1}-k_{1}^{0}\vec{k}_{2}\right| }%
\left( \prod_{f=1}^{2}\frac{d^{3}p_{f}}{\left( 2\pi \right) ^{3}2E_{f}}%
\right) \left| M\right| ^{2}\left( 2\pi \right) ^{4}
\end{eqnarray*}
and 
\[
\left| M\right| ^{2}=g_{s}^{2}O_{ij}A_{ij}=g_{s}^{2}\left(
O_{11}A_{11}+O_{22}A_{22}+O_{c}\left( A_{12}+A_{21}\right) \right) , 
\]
where 
\begin{eqnarray*}
O_{11} &=&\frac{1}{4\left( k_{1}\cdot p_{1}\right) ^{2}}, \\
O_{22} &=&\frac{1}{4\left( k_{1}\cdot p_{2}\right) ^{2}}, \\
O_{c} &=&O_{12}=O_{21}=\frac{1}{4\left( k_{1}\cdot p_{1}\right) \left(
k_{1}\cdot p_{2}\right) },
\end{eqnarray*}
and after averaging over gluon colours and transverse polarizations, and
summing over fermion colours we have 
\begin{eqnarray*}
A_{11} &=&A_{11}^{\left( +\right) }+A_{11}^{\left( -\right)
}+m_{t}m_{b}\varepsilon ^{2}\frac{g_{R}^{\ast }g_{L}+g_{L}^{\ast }g_{R}}{2}%
\left\{ m_{t}^{2}-k_{1}\cdot p_{1}\right\} \\
&&+m_{t}^{2}\left( \left| g_{L}\right| ^{2}+\left| g_{R}\right| ^{2}\right)
\left\{ 2\left( \varepsilon \cdot p_{2}\right) \left( \varepsilon \cdot
\left( k_{1}-p_{1}\right) \right) -\varepsilon ^{2}\left( p_{2}\cdot \left(
k_{1}-p_{1}\right) \right) \right\} ,
\end{eqnarray*}
with\ 
\begin{eqnarray*}
A_{11}^{\left( \pm \right) } &\equiv &\left| g_{\pm }\right| ^{2}\varepsilon
\cdot p_{2}\left\{ 2\left( \left( k_{1}-p_{1}\right) \cdot \varepsilon
\right) \left( \left( k_{1}-p_{1}\right) \cdot \frac{p_{1}\pm m_{t}n}{2}%
\right) -\left( k_{1}-p_{1}\right) ^{2}\left( \varepsilon \cdot \frac{%
p_{1}\pm m_{t}n}{2}\right) \right\} \\
&&-\left| g_{\pm }\right| ^{2}\frac{\varepsilon ^{2}}{2}\left\{ 2\left(
\left( k_{1}-p_{1}\right) \cdot p_{2}\right) \left( \left(
k_{1}-p_{1}\right) \cdot \frac{p_{1}\pm m_{t}n}{2}\right) -\left(
k_{1}-p_{1}\right) ^{2}\left( p_{2}\cdot \frac{p_{1}\pm m_{t}n}{2}\right)
\right\} \\
&&+\left| g_{\pm }\right| ^{2}\frac{m_{t}^{2}}{2}\left\{ 2\left( \varepsilon
\cdot p_{2}\right) \left( \varepsilon \cdot \frac{p_{1}\mp m_{t}n}{2}\right)
-\varepsilon ^{2}\left( p_{2}\cdot \frac{p_{1}\mp m_{t}n}{2}\right) \right\}
,
\end{eqnarray*}
and 
\[
A_{22}=A_{22}^{\left( +\right) }+A_{22}^{\left( -\right) }+m_{b}m_{t}\frac{%
g_{R}^{\ast }g_{L}+g_{L}^{\ast }g_{R}}{2}\left\{ p_{2}\cdot \left(
p_{2}-k_{1}\right) \right\} \varepsilon ^{2}, 
\]
with 
\begin{eqnarray*}
A_{22}^{\left( \pm \right) } &\equiv &\left| g_{\pm }\right| ^{2}p_{2}\cdot
\left( k_{2}-p_{1}\right) \left\{ 2\left( \varepsilon \cdot \left(
k_{2}-p_{1}\right) \right) \left( \varepsilon \cdot \frac{p_{1}\pm m_{t}n}{2}%
\right) -\varepsilon ^{2}\left( \left( k_{2}-p_{1}\right) \cdot \frac{%
p_{1}\pm m_{t}n}{2}\right) \right\} \\
&&-\frac{\left| g_{\pm }\right| ^{2}}{2}\left( k_{2}-p_{1}\right)
^{2}\left\{ 2\left( \varepsilon \cdot p_{2}\right) \left( \varepsilon \cdot 
\frac{p_{1}\pm m_{t}n}{2}\right) -\varepsilon ^{2}\left( p_{2}\cdot \frac{%
p_{1}\pm m_{t}n}{2}\right) \right\} \\
&&+m_{b}^{2}\frac{\left| g_{\pm }\right| ^{2}}{2}\left\{ 2\left( \varepsilon
\cdot \left( 4k_{1}-3p_{2}\right) \right) \left( \varepsilon \cdot \frac{%
p_{1}\pm m_{t}n}{2}\right) -\varepsilon ^{2}\left( \left(
4k_{1}-3p_{2}\right) \cdot \frac{p_{1}\pm m_{t}n}{2}\right) \right\}
\end{eqnarray*}
and 
\begin{eqnarray*}
A_{12}+A_{21} &=&A_{c}^{\left( +\right) }+A_{c}^{\left( -\right) } \\
&&-m_{b}m_{t}\frac{g_{R}^{\ast }g_{L}+g_{L}^{\ast }g_{R}}{2}\left\{
\varepsilon ^{2}\left[ 2\left( p_{2}\cdot p_{1}\right) -\left( \left(
p_{2}+p_{1}\right) \cdot k_{1}\right) \right] +2\left( \varepsilon \cdot
k_{1}\right) ^{2}\right\} \\
&&+\frac{\left| g_{R}\right| ^{2}+\left| g_{L}\right| ^{2}}{2}\left\{
2m_{t}^{2}\left( \varepsilon \cdot p_{2}\right) \left( \varepsilon \cdot
\left( k_{2}-p_{1}\right) \right) -m_{b}^{2}m_{t}^{2}\varepsilon
^{2}\right\} ,
\end{eqnarray*}
with 
\begin{eqnarray*}
A_{c}^{\left( \pm \right) } &=&2\left| g_{\pm }\right| ^{2}\left(
\varepsilon \cdot p_{2}\right) \left\{ \left( \left( k_{1}-p_{1}\right)
\cdot \frac{p_{1}\pm m_{t}n}{2}\right) \left( \varepsilon \cdot \left(
k_{2}-p_{1}\right) \right) \right. \\
&&-\left. \left( \frac{p_{1}\pm m_{t}n}{2}\cdot \left( k_{2}-p_{1}\right)
\right) \left( k_{1}\cdot \varepsilon \right) \right\} \\
&&-2\left| g_{\pm }\right| ^{2}\left( \varepsilon \cdot \frac{p_{1}\pm m_{t}n%
}{2}\right) \left\{ \left( \varepsilon \cdot p_{2}\right) \left( p_{1}\cdot
\left( k_{1}-p_{2}\right) \right) \right. \\
&&+\left( \varepsilon \cdot \left( k_{2}-p_{1}\right) \right) \left( \left(
k_{1}-p_{1}\right) \cdot p_{2}\right) \\
&&-\left. \left( \varepsilon \cdot \left( k_{1}-p_{1}\right) \right) \left(
p_{2}\cdot \left( k_{2}-p_{1}\right) \right) \right\} \\
&&+\left| g_{\pm }\right| ^{2}\varepsilon ^{2}\left\{ \left( p_{2}\cdot
\left( k_{1}-p_{1}\right) \right) \left( \left( k_{2}-p_{1}\right) \cdot 
\frac{p_{1}\pm m_{t}n}{2}\right) \right. \\
&&+\left( p_{2}\cdot \frac{p_{1}\pm m_{t}n}{2}\right) \left( \left(
k_{2}-p_{1}\right) \cdot \left( k_{1}-p_{1}\right) \right) \\
&&-\left. \left( p_{2}\cdot \left( k_{2}-p_{1}\right) \right) \left( \left(
k_{1}-p_{1}\right) \cdot \frac{p_{1}\pm m_{t}n}{2}\right) \right\} \\
&&-2\left| g_{\pm }\right| ^{2}m_{b}^{2}\left( \varepsilon \cdot \left(
k_{1}-p_{1}\right) \right) \left( \varepsilon \cdot \frac{p_{1}\pm m_{t}n}{2}%
\right) ,
\end{eqnarray*}
where $\varepsilon $ is the polarization of the $W^{+}$ boson and 
\begin{eqnarray*}
g_{+} &\equiv &g_{R}, \\
g_{-} &\equiv &g_{L},
\end{eqnarray*}
For the subprocess of Fig.\ref{t-b-} we have to perform the following
changes in the expressions for the single top production. 
\begin{eqnarray*}
\varepsilon &\leftrightarrow &\varepsilon ^{\ast }, \\
n &\leftrightarrow &-n, \\
p_{2} &\leftrightarrow &-p_{2}, \\
p_{1} &\leftrightarrow &-p_{1}, \\
k_{2} &\leftrightarrow &-k_{2}, \\
k_{1} &\leftrightarrow &-k_{1},
\end{eqnarray*}
but since we can take the W-boson polarization real and the cross section is
even under the above sign changes, the subprocess cross section is the same
for single top or anti-top production.

\end{document}